\documentstyle[12pt]{article}
\def\e{{\rm e}}
\newcommand{\be}{\begin{equation}}
\newcommand{\ee}{\end{equation}}
\newcommand{\bea}{\begin{eqnarray}}
\newcommand{\eea}{\end{eqnarray}}
\newcommand{\al}{\alpha}

\newcommand{\gm}{\gamma}
\newcommand{\Gm}{\Gamma}
\newcommand{\dl}{\delta}

\newcommand{\ep}{\epsilon}

\newcommand{\lm}{\lambda}

\newcommand{\dd}{\mbox{d}}

\newcommand{\nn}{\nonumber}
\newcommand{\Li}[2]{{\mbox{Li}}_{#1}\left(#2\right)}

\begin{document}
\parindent=1.5pc

\begin{titlepage}
\rightline{}
\rightline{hep-ph/9905323}
\rightline{May 1999}
\bigskip
\begin{center}
{{\bf
Analytical Result for Dimensionally Regularized \\
Massless On-shell Double Box
} \\
\vglue 5pt
\vglue 1.0cm
{ {\large V.A. Smirnov\footnote{E-mail: smirnov@theory.npi.msu.su.}
} }\\
\baselineskip=14pt
\vspace{2mm}
{\em Nuclear Physics Institute of Moscow State University}\\
{\em Moscow 119899, Russia}
\vglue 0.8cm
{Abstract}}
\end{center}
\vglue 0.3cm
{\rightskip=3pc
 \leftskip=3pc
\noindent
The dimensionally regularized massless on-shell double box Feynman
diagram with powers of propagators equal to one is analytically
evaluated for general values of the
Mandelstam variables $s$ and $t$. An explicit result is expressed
either in terms
of polylogarithms $\Li{a}{ -t/s }$, up to $a=4$, and generalized
polylogarithms $S_{a,b}(-t/s)$, with $a=1,2$ and $b=2$,
or in terms of these functions depending on the inverse ratio, $s/t$.
\vglue 0.8cm}
\end{titlepage}

\section{Introduction}

The massless double box diagram shown in Fig.~1 enters many important
physical observables, e.g., amplitudes of the Bhabba scattering
at high energies.
An experience shows that master diagrams, i.e. with all powers of
propagators equal to one, are most complicated for evaluation.
In the massless off-shell case,
the master double box Feynman integral
has been analytically evaluated in \cite{UD} strictly in four dimensions.
It is the purpose of the present paper to evaluate it analytically
on shell, i.e. for $p_i^2=0,\; i=1,2,3,4$,
in the framework of dimensional regularization \cite{dimreg},
with the space-time dimension $d=4-2\ep$ as a regularization parameter.
\begin{figure}[hbt]
\centering
\begin{picture}(160,60)(0,2)
\put(20,10){\line(1,0){120}}
\put(20,50){\line(1,0){120}}
\put(40,10){\line(0,1){40}}
\put(80,10){\line(0,1){40}}
\put(120,10){\line(0,1){40}}
\put(40,10){\circle*{3}}
\put(40,50){\circle*{3}}
\put(120,10){\circle*{3}}
\put(120,50){\circle*{3}}
\put(80,10){\circle*{3}}
\put(80,50){\circle*{3}}
\put(6,9){$p_1$}
\put(6,49){$p_2$}
\put(146,9){$p_3$}
\put(146,49){$p_4$}
\put(32,27){\small $5$}
\put(72,27){\small $6$}
\put(112,27){\small $7$}
\put(59,0){\small $3$}
\put(99,0){\small $1$}
\put(59,53){\small $4$}
\put(99,53){\small $2$}
\end{picture}
\caption{}
\end{figure}

To do this, we start, in the next Section, from the
alpha-representation of the double box and, after expanding some
of the involved functions
in Mellin--Barnes (MB) integrals, arrive at a five-fold
MB integral representation with gamma functions in the integrand.
We use, in Sec.~3, a standard procedure of taking residues and shifting
contours to resolve the structure of singularities in
the parameter of dimensional regularization, $\ep$.
This procedure leads to an appearance of multiple terms where
Laurent expansion
in $\ep$ becomes possible. The resulting integrals in all the MB parameters
but one are evaluated explicitly in gamma functions and their derivatives.
In Sec.~4, the last MB integral is evaluated by closing an initial
integration contour in the complex plane to the right,
with an explicit summation of the corresponding series.
A final  result is expressed in terms of
polylogarithms $\Li{a}{ -t/s }$, up to $a=4$, and generalized
polylogarithms $S_{a,b}(-t/s)$, with $a=1,2$ and $b=2$.
Starting from the same one-fold MB integral and closing the contour
of integration to the left, we obtain a similar result written
through the same class of functions depending on the
inverse ratio, $s/t$. Furthermore, we obtain, as a by-product,
an explicit result for the backward scattering value, i.e. at $t=-s$,
of the double box diagram.

\section{From momentum space to MB representation}

The massless on-shell double box Feynman integral can be written as
\bea
\int\int \frac{\dd^dk \dd^dl}{(k^2+2 p_1 k) (k^2-2 p_2 k)
k^2 (k-l)^2
(l^2+2 p_1 l)(l^2-2 p_2 l) (l-p_1-p_3)^2 }
\nn \\
\equiv
\frac{\left(i\pi^{d/2} \e^{-\gm_{\rm E}\ep} \right)^2 }{(-s)^{2+2\ep}(-t)}
K(t/s,\ep)
\, ,
\label{2box}
\eea
where $s=(p_1+p_2)^2, \;  t=(p_1+p_3)^2$, and
$k$ and $l$ are respectively loop momenta of the left and the right box.
Usual prescriptions, $k^2=k^2+i 0, \; -s=-s-i 0$, etc are implied.
We have pulled out not only standard factors that arise when integrating
in the loop momenta but also a factor that makes the resulting
function $K$ depend on the dimensionless variable, $x=t/s$.

The alpha representation of the double box is straightforwardly obtained:
\be
K(x,\ep) = -\Gm(3+2\ep) \int_0^\infty \dd\al_1 \ldots\int_0^\infty\dd\al_7
\dl\left( \sum \al_i-1\right) D^{1+3\ep}
\left(A+x\al_5\al_6\al_7\right)^{-3-2\ep} \; ,
\label{alpha}
\ee
where
\bea
D&=&(\al_1+\al_2+\al_7) (\al_3+\al_4+\al_5)
+\al_6 (\al_1+\al_2+\al_3+\al_4+\al_5+\al_7) \;, \\
A&=& \al_1\al_2 (\al_3+\al_4+\al_5) + \al_3\al_4(\al_1+\al_2+\al_7)
+\al_6 (\al_1+\al_3)( \al_2+\al_4) \; .
\eea
As it is well-known, one can choose
a sum of an arbitrary subset of $\al_i\,, i=1,\ldots,7$ in
the argument of the delta function in (\ref{alpha}).
We  choose it as
$\dl\left( \sum_{i\neq 6} \al_i-1\right)$ and change variables by turning
from alpha to Feynman parameters
\bea
\al_3 = \al_{35} \xi_1,\; \al_5 = \al_{35}  (1-\xi_1), \;
\al_1 = \al_{17} \xi_3, \; \al_7 = \al_{17}  (1-\xi_3),\;
\nn \\
\al_{35} = \xi_5 \xi_2, \; \al_4 = \xi_5 (1-\xi_2), \;
\al_{17} = (1-\xi_5) \xi_4, \;  \al_2 =(1-\xi_5) (1-\xi_4).
\eea
to obtain the following parametric integral:
\bea
K(x,\ep) = -\Gm(3+2\ep) \int_0^\infty \dd\al_6
\int_0^1\dd\xi_1\ldots\int_0^1\dd\xi_5 \;
\xi_2 \xi_4 \xi^2 (1-\xi)^2
\nn \\ && \hspace*{-60mm}\times
(\al_6+\xi_5 (1-\xi_5))^{1+3\ep} Q^{-3-2\ep} \; ,
\label{Feynman}
\eea
where
\bea
Q &= & x \al_6 (1-\xi_1) \xi_2 (1-\xi_3) \xi_4 (1-\xi_5) \xi_5
\nn \\ && \hspace*{-5mm}
+\xi_5 (1-\xi_5) [\xi_5 \xi_1 \xi_2 (1-\xi_2) +
(1-\xi_5) \xi_3 \xi_4 (1-\xi_4)]
\nn \\ && \hspace*{-5mm}
+\al_6 [\xi_5 \xi_1 \xi_2 +(1-\xi_5) \xi_3 \xi_4 ] [\xi_5 (1- \xi_2)
+(1-\xi_5) (1-\xi_4) ] \; .
\label{Q}
\eea

We are now going to apply five times the MB representation
\be
\frac{1}{(X+Y)^{\nu}} = \frac{1}{\Gm(\nu)}
\frac{1}{2\pi i}\int_{-i \infty}^{+i \infty} \dd w
\frac{Y^w}{X^{\nu+w}} \Gm(\nu+w) \Gm(-w) \; ,
\label{MB}
\ee
where the contour of integration is chosen in the standard way:
the poles with the $\Gm(\ldots+w)$-dependence
(let us call them infrared (IR) poles) are to the left of the contour and
the poles with the $\Gm(\ldots-w)$-dependence (ultraviolet (UV) poles) are
to the right of it.

\mbox{}First, we introduce a MB integration, in $w$, using decomposition of
the function $Q$ with $Y$ as the first line in (\ref{Q}).
We introduce a second MB integral choosing as $X$ the term
with $\al_6$ in the rest part of $Q$.
After that we can take an integral in $\al_6$ in gamma functions.
The next three MB integrations, in $z_1,z_2$ and $z_3$,
are to separate terms in the
following three combinations:
$[\xi_5 \xi_1 \xi_2 +(1-\xi_5) \xi_3 \xi_4 ]$,
$[\xi_5 (1- \xi_2) +(1-\xi_5) (1-\xi_4) ]$ and
$[\xi_5 \xi_1 \xi_2 (1-\xi_2) + (1-\xi_5) \xi_3 \xi_4 (1-\xi_4)]$.

All the integrals in Feynman parameters are then taken explicitly
in gamma functions. Finally, we perform the change of
variables $z_2=w_2-z_1-1,\; z_3=w_3-z_1-1$
and arrive at the following nice 5-fold MB integral:
\bea
K(x,\ep) =
- \frac{1}{\Gm(-1-3 \ep)} \frac{1}{(2\pi i)^5}
\int \dd w \dd w_2\dd w_3 \dd z\dd z_1 x^{w+1}
\nn \\ \times
\Gm(1+w)^2 \Gm(-w)
\Gm(w_2) \Gm(-1-2 \ep-w-w_2)\Gm(w_3) \Gm(-1-2 \ep-w-w_3)
\nn \\ \times
\frac{\Gm(1-w_2+z_1) \Gm(1-w_3+z_1) \Gm(\ep+w+w_2+w_3-z_1) \Gm(-z_1)}
{ \Gm(1+w+w_2+w_3) \Gm(-1-4 \ep-w-w_2-w_3) }
\nn \\ \times
\Gm(1-\ep+z)
\Gm(2+2 \ep+w+w_2+z-z_1) \Gm(2+2 \ep+w+w_3+z-z_1)
\nn \\ \times
\frac{\Gm(-2-3 \ep-w-w_2-w_3+z_1-z) \Gm(z_1-z)}{ \Gm(3+2\ep+w+z)} \, .
\label{5MB}
\eea
One can interchange the order of integration in an arbitrary way. For
each order, the rules of dealing with poles are as formulated above.
Note that if we have a product $\Gm(a+v) \Gm(b-v)$, for some integration
variable $v=w,w_2,w_3,z,z_1$ with $a$ and $b$ dependent
on other variables, then the integration in $v$ produces a singularity
of the type $\Gm(a+b)$.

\section{Resolving singularities in $\ep$}

Since it looks hopeless to evaluate our MB integral for general $\ep$
let us try to obtain a result in expansion in $\ep$ up to the finite part.
There is a factor $1/\Gm(-1-3 \ep)$ proportional to $\ep$
when $\ep$ tends to zero. Representation (\ref{5MB}) is therefore
effectively 4-fold
because to generate a contribution that does not vanish at $\ep=0$
we need to take a residue at least in one of the integration variables.
None of the integrations can however immediately produce an
explicit $\ep$-pole.
Let us first distinguish the following two gamma functions
\[ \Gm(\ep+w+w_2+w_3-z_1) \; \Gm(-2-3 \ep-w-w_2-w_3+z_1-z) \]
that are essential for the appearance of the poles.

We can write down the integral in $z_1$ as minus residue at the point
$z_1=\ep+w+w_2+w_3$ (where the gamma function $\Gm(\ep+w+w_2+w_3-z_1)$
has its first pole which is UV, with respect to $z_1$) plus an
integral with the same integrand where this pole is IR.
We can similarly write down the integral in $z$ as minus residue at the
point $z=-2-3 \ep-w-w_2-w_3+z_1$
(where the gamma function $\Gm(-2-3 \ep-w-w_2-w_3+z_1-z)$
has its first pole which is UV, with respect to $z$) plus an
integral with the same integrand where this pole is IR.

As a result we decompose integral (\ref{5MB}) as
$K=K_{00}+K_{01}+K_{10}+K_{11}$ where $K_{11}$ corresponds to the two
residues, $K_{10}$ to the residue in $z$ and the integral in $z_1$
with the opposite nature of the first pole of $\Gm(\ep+w+w_2+w_3-z_1)$,
etc. For example, the contribution $K_{11}$ is given by the
following 3-fold integral:
\bea
K_{11}(x,\ep) =
- \frac{1}{(2\pi i)^3}
\int \dd w \dd w_2\dd w_3 x^{w+1}
\Gm(1+w) \Gm(-w)
\nn \\ \times
\Gm(w_2) \Gm(-\ep-w_2) \Gm(1+\ep+w+w_2) \Gm(-1-2 \ep-w-w_2)
\nn \\ \times
\Gm(w_3) \Gm(-\ep-w_3) \Gm(1+\ep+w+w_3) \Gm(-1-2 \ep-w-w_3)
\nn \\ \times
\frac{\Gm(2+3\ep +w+w_2+w_3) \Gm(-\ep -w-w_2-w_3)}
{ \Gm(1+w+w_2+w_3) \Gm(-1-4 \ep-w-w_2-w_3) } \, .
\label{K11}
\eea
This contribution is in turn decomposed, in a similar way,
as $K_{11}=\sum_{i,j=0,1,2} K_{11ij}$. Here the value $i=1$ of the
first index denotes the residue in $w_2$ at the point $w_2=0$. The value
$i=2$ denotes the residue in $w_2$ at $w_2=-1-\ep-w$ of the integrand
where the first pole of $\Gm(w_2)$ is UV rather than IR. Finally, $i=0$
means that both above poles are IR. The second index similarly refers
to the integral in $w_3$.

In particular, we have
\bea
K_{1111}(x,\ep) = -\Gm(-\ep)^2 \frac{1}{2\pi i}
\int \dd w  x^{w+1}
\nn \\ \times
\frac{\Gm(1+\ep+w)^2 \Gm(2+3\ep+w) \Gm(-1-2\ep-w)^2  \Gm(-\ep-w)
\Gm(-w)}{\Gm(-1-4\ep-w)} \, ,
\label{K1111} \\
K_{1112}(x,\ep) = K_{1121}(x,\ep) =
\frac{\Gm(1+2\ep) \Gm(-\ep)}{\Gm(-3\ep)}
\frac{1}{2\pi i}
\int \dd w  x^{w+1}
\nn \\ \times
\frac{\Gm(1+w)^2 \Gm(1+\ep+w)^2 \Gm(-1-2\ep-w)  \Gm(-\ep-w)
\Gm(-w)}{\Gm(2+\ep+w)} \, , \label{K1112} \\
K_{1122}(x,\ep) = -\Gm(-\ep)^2 \frac{1}{2\pi i}
\int \dd w  x^{w+1}
\nn \\ \times
\frac{\Gm(1+w)^3 \Gm(1+\ep+w)^2 \Gm(-\ep-w)^2  \Gm(\ep-w)
\Gm(-w)}{\Gm(2+\ep+w) \Gm(1-2\ep+w) \Gm(-1-2\ep-w) } \, .
\label{K1122}
\eea

The next contribution is
\bea
K_{10}(x,\ep) =
- \frac{1}{\Gm(-1-3 \ep)} \frac{1}{(2\pi i)^4}
\int \dd w \dd w_2\dd w_3 \dd z_1 x^{w+1}
\Gm(1+w)^2 \Gm(-w)
\nn \\ \times
\Gm(w_2) \Gm(-\ep-w_2) \Gm(-1-2 \ep-w-w_2)
\Gm(w_3) \Gm(-\ep-w_3) \Gm(-1-2 \ep-w-w_3)
\nn \\ \times
\frac{\Gm(2+3 \ep+w+w_2+w_3) \Gm(1-w_2+z_1) \Gm(1-w_3+z_1) }{
\Gm(1+w+w_2+w_3) \Gm(-1-4 \ep-w-w_2-w_3)}
\nn \\ \times
\frac{
\Gm(-1-4 \ep-w-w_2-w_3+z_1) \Gm(\ep+w+w_2+w_3-z_1) \Gm(-z_1)}
{  \Gm(1-\ep-w_2-w_3+z_1) }
\, ,
\label{K10}
\eea
where the first pole of $\Gm(\ep+w+w_2+w_3-z_1)$ is IR, with respect
to $z_1$, rather than UV. We further decompose this contribution by changing
the nature of the first pole of $\Gm(-1-4 \ep-w-w_2-w_3+z_1)$ in $z_1$.
We obtain $K_{10}=K_{100}+K_{101}$, where
the new index 1 corresponds to the residue and has the form
\bea
K_{101}(x,\ep) =
- \frac{1}{(2\pi i)^3}
\int \dd w \dd w_2\dd w_3 x^{w+1}
\frac{\Gm(1+w)^2 \Gm(-w)}{\Gm(2+3\ep+w)}
\nn \\ \times
\Gm(w_2) \Gm(-\ep-w_2) \Gm(2+4\ep+w+w_2) \Gm(-1-2 \ep-w-w_2) \Gm(w_3)
\nn \\ \times
\frac{ \Gm(-\ep-w_3) \Gm(2+4\ep+w+w_3) \Gm(-1-2 \ep-w-w_3)
\Gm(2+3 \ep+w+w_2+w_3)}{\Gm(1+w+w_2+w_3)} \, .
\label{K101}
\eea
Each of the contributions $K_{100}$ and $K_{101}$ is
then decomposed using the change of the nature of poles $w_2=0$ and $w_3=0$.
We obtain
$K_{10j}=K_{10j00}+K_{10j01}+K_{10j10}+K_{10j11}$, for $j=0,1$.
Here the value $i=1$ of the
last index denotes the residue in $w_3$ at the point $w_3=0$ and
the $i=0$ an integral where the first pole of $\Gm(w_3)$ is considered UV.
The second index from the end similarly refers to $\Gm(w_2)$. For example,
\be
K_{10111}(x,\ep) = -\Gm(-\ep)^2 \frac{1}{2\pi i}
\int \dd w  x^{w+1}
\Gm(2+4\ep+w)^2 \Gm(1+w) \Gm(-1-2\ep-w)^2 \Gm(-w) \,  .
\label{K0111}
\ee

Then we have
\bea
K_{01}(x,\ep) =
- \frac{1}{\Gm(-1-3 \ep)} \frac{1}{(2\pi i)^4}
\int \dd w \dd w_2\dd w_3 \dd z x^{w+1}
\Gm(1+w)^2 \Gm(-w)
\nn \\ \times
\Gm(w_2) \Gm(1+\ep+w+w_2) \Gm(-1-2 \ep-w-w_2)
\Gm(w_3) \Gm(1+\ep+w+w_3)
\nn \\ \times
\frac{\Gm(-1-2 \ep-w-w_3) \Gm(-\ep-w-w_2-w_3) \Gm(1-\ep+z) }{
\Gm(1+w+w_2+w_3) \Gm(-1-4 \ep-w-w_2-w_3)}
\nn \\ \times
\frac{
\Gm(2+\ep-w_2+z) \Gm(2+\ep-w_3+z) \Gm(\ep+w+w_2+w_3-z) \Gm(-2-2\ep-z)}
{  \Gm(3+2\ep+w+z) }
\, ,
\label{K01}
\eea
where the first pole of $\Gm(-2-2\ep-z)$ is IR, with respect to $z$,
rather than UV.
Using the change of variables
$w_2\to -1-2 \ep -w-w_2, \, w_3\to -1-2 \ep -w-w_3,
z_1\to -\ep-w-w_2-w_3+z$
in $K_{10}$ we see that $K_{01}\equiv K_{10}$. Finally,
the contribution $K_{00}$ is similarly decomposed:
$K_{00}=K_{0000}+K_{0001}+K_{0010}+K_{0011}$.

Now we observe that, in each of the obtained contributions,
the only additional (with respect to explicit
gamma functions depending on $\ep$) source of the poles in $\ep$
is the last integration, in $w$, where the first (UV) pole of the
gamma function $\Gm(-1-2 \ep-w)$ glues with an IR pole of $\Gm(1+w)$ or
$\Gm(1+\ep+w)$ when $\ep\to 0$
--- see such examples in (\ref{K1111}--\ref{K1122}) and (\ref{K0111}).
Therefore we further decompose each of the contributions
into two pieces: minus residue at the point $w=-1-2 \ep$ plus
an integral where we can integrate in the region $-1<{\rm Re}w<0$.
In each of these pieces, we now can expand an integrand in a Laurent series
in $\ep$ up to the finite part.
In particular, no poles in $\ep$ arise in $K_{0000}$ so that it it zero
at $\ep=0$ because of the overall factor $1/\Gm(-1-3 \ep)$.

We collect separately the pieces from these last residues and from the last
integration at $-1<{\rm Re}w<0$. The first collection gives the leading
order term in the expansion of the double box in the limit $t/s\to 0$ while
the second collection involves the rest of the terms of this expansion.
A remarkable fact is that, in all these multiple contributions,
the integrations in $w_2,w_3,z,z_1$
can be performed analytically, with the help of
the first and the second Barnes lemmas
\bea
\frac{1}{2\pi i}\int_{-i \infty}^{+i \infty} \dd w \,
\Gm(\lm_1+w) \Gm(\lm_2+w) \Gm(\lm_3-w) \Gm(\lm_4-w)
& &
\nn \\ && \hspace*{-70mm} =
\frac{\Gm(\lm_1+\lm_3) \Gm(\lm_1+\lm_4) \Gm(\lm_2+\lm_3)
\Gm(\lm_2+\lm_4)}{\Gm(\lm_1+\lm_2+\lm_3+\lm_4)} \; ,
\label{1stBarnes}
\\
\frac{1}{2\pi i}\int_{-i \infty}^{+i \infty} \dd w \,
\frac{\Gm(\lm_1+w) \Gm(\lm_2+w) \Gm(\lm_3+w) \Gm(\lm_4-w) \Gm(\lm_5-w)}{
\Gm(\lm_1+\lm_2+\lm_3+\lm_4+\lm_5+w) }
&  &
\nn \\ && \hspace*{-110mm} =
\frac{\Gm(\lm_1+\lm_4) \Gm(\lm_2+\lm_4) \Gm(\lm_3+\lm_4)
      \Gm(\lm_1+\lm_5) \Gm(\lm_2+\lm_5) \Gm(\lm_3+\lm_5)
}{\Gm(\lm_1+\lm_2+\lm_4+\lm_5) \Gm(\lm_1+\lm_3+\lm_4+\lm_5)
\Gm(\lm_2+\lm_3+\lm_4+\lm_5)} \;
\label{2ndBarnes}
\eea
and their corollaries. These are two typical examples of such corollaries:
\bea
\frac{1}{2\pi i}\int_{-i \infty}^{+i \infty} \dd w \,
\frac{\Gm(\lm_1+w) \Gm(\lm_2+w)^2 \Gm(-\lm_2-w) \Gm(\lm_3-w)}{
\Gm(\lm_1+\lm_2+\lm_3+w) }
&  &
\nn \\ && \hspace*{-70mm} =
\frac{
\Gm(\lm_1-\lm_2) \Gm(\lm_2+\lm_3)
\left[
\psi'\left(\lm_1+\lm_3\right) - \psi'\left(\lm_2+\lm_3\right)\right]}{
\Gm(\lm_1+\lm_3)} \; ,
\label{2ndBex}
\eea
where the pole  $w=-\lm_2$ is considered IR
while other poles are treated in the standard way, and
\be
\frac{1}{2\pi i}\int_{-1/2-i \infty}^{-1/2+i \infty} \dd w \,
\Gm(1+w) \Gm(w) \Gm(-w) \Gm(-1-w) \psi(1+w)^2
=\frac{\gm_{\rm E}^2 \pi^2}{3} +6\gm_{\rm E}\zeta(3) +\frac{\pi^4}{45}
\; .
\ee
Here $\gm_{\rm E}$ is the Euler constant, $\psi(z)$ the logarithmical
derivative of the gamma function, and $\zeta(z)$ the Riemann zeta function.

After taking these integrations and summing up the resulting contributions
into the two above collections we obtain the following result
\bea
K(x,\ep) &= & K_{0t}(x,\ep)+ K_{1t}(x,\ep) + o(\ep) \; , \\
K_{0t}(x,\ep) & = &
-\frac{4}{\ep^4} +\frac{5\ln x}{\ep^3}
- \left( 2 \ln^2 x -\frac{5}{2} \pi^2  \right) \frac{1}{\ep^2}
\nn \\ && \hspace*{-30mm}
-\left( \frac{2}{3}\ln^3 x +\frac{11}{2}\pi^2 \ln x
-\frac{65}{3} \zeta(3) \right) \frac{1}{\ep}
+\frac{4}{3}\ln^4 x +6 \pi^2 \ln^2 x
-\frac{88}{3} \zeta(3)\ln x +\frac{29}{30}\pi^4 \; ,
\label{K0t}
\\
K_{1t}(x,\ep) & = &
\frac{2}{\pi i} \int \frac{\dd w x^{w+1}}{1+w}
\Gm(1+w)^3 \Gm(-w)^3
\nn \\ && 
\times
\left[
\frac{1}{\ep} -\frac{5}{1+w} + 3 \psi(1+w)
-4 \psi(-w) -\gm_{\rm E} \right] \; .
\label{K1tMB}
\eea

Let us stop for a moment and observe that this result provides,
in a very easy way,
not only numerical evaluation of the double box diagram
for general values of $s$ and $t$ but also
asymptotic expansions in the limits $t/s\to 0$ and $s/t\to 0$
which are obtained by taking series of residues respectively
to the right or to the left.

\section{Evaluating the last MB integral}

The last MB integration, in (\ref{K1tMB}), is performed analytically
by taking the sum of the residues at the points $w=0,1,2,\ldots$
and summing up the resulting series. In this last step, we use,
in particular, summation formulae derived in \cite{Oleg}.
Here is the final result:
\bea
K_{1t}(x,\ep) & = &
- \left[
2 \Li{3}{ -x } -2\ln x \Li{2}{ -x }
-\left( \ln^2 x +\pi^2 \right) \ln(1+x)
\right] \frac{2}{\ep}
\nn \\ && \hspace*{-30mm}
- 4 \left(S_{2,2}(-x) - \ln x S_{1,2}(-x)  \right)
+ 44 \Li{4}{ -x } - 4 \left(\ln(1+x) + 6 \ln x  \right) \Li{3}{ -x }
\nn \\ && \hspace*{-30mm}
+ 2\left( \ln^2 x +2 \ln x \ln(1+x) +\frac{10}{3}\pi^2\right) \Li{2}{-x}
\nn \\ && \hspace*{-30mm}
+\left( \ln^2 x +\pi^2 \right) \ln^2(1+x)
-\frac{2}{3} \left(4\ln^3 x +5\pi^2 \ln x -6\zeta(3)\right) \ln(1+x)
\; ,
\eea
where $\Li{a}{z}$ is the polylogarithm \cite{Lewin} and
\be
\label{Sab}
  S_{a,b}(z) = \frac{(-1)^{a+b-1}}{(a-1)! b!}
    \int_0^1 \frac{\ln^{a-1}(t)\ln^b(1-zt)}{t} \dd t \; ,
\ee
a generalized polylogarithm introduced in \cite{Devoto}.
Note that any (generalized) polylogarithms involved can be expanded
in a Taylor series at $x=0$ with the radius of convergence equal to one.

We can similarly close the integration contour to the left and obtain
a result in a form of functions depending on the inverse ratio, $y=1/x$:
\bea
K(x,\ep) &= &  K_{0s}(x,\ep)+ K_{1s}(x,\ep) + o(\ep) \; , \\
K_{0s}(1/y,\ep) & = &
-\frac{4}{\ep^4} -\frac{5\ln y}{\ep^3}
- \left( 2 \ln^2 y -\frac{5}{2} \pi^2  \right) \frac{1}{\ep^2}
\nn \\ && \hspace*{-30mm}
+ \left( \frac{7}{2}\pi^2\ln y
+ \frac{65}{3} \zeta(3) \right) \frac{1}{\ep}
+ \frac{1}{3} \pi^2 \ln^2 y
+\frac{76}{3} \zeta(3)\ln y -\frac{83}{90}\pi^4 \; ,
\\
K_{1s}(1/y,\ep) & =&
-\left[
2 \Li{3}{ -y } -2\ln y \Li{2}{ -y }
-\left( \ln^2 y +\pi^2 \right) \ln(1+y)
\right] \frac{2}{\ep}
\nn \\ && \hspace*{-30mm}
- 4 \left( S_{2,2}(-y) - \ln y S_{1,2}(-y)  \right)
-36 \Li{4}{ -y } - 4 \left(\ln(1+y) - 5 \ln y \right)
\Li{3}{ -y }
\nn \\&& \hspace*{-30mm}
-2 \left( \ln^2 y - 2 \ln y \ln(1+y) +
\frac{10}{3}\pi^2\right) \Li{2}{ -y }
\nn \\ && \hspace*{-30mm}
+\left( \ln^2 y +\pi^2 \right) \ln^2(1+y)
+2 \left(\ln^3 y +\frac{2}{3}\pi^2 \ln y +2\zeta(3)\right)
\ln(1+y)
\; .
\eea

As a by-product, we obtain an explicit result
for the backward scattering value of (\ref{2box}), i.e. at $t=-s$,
\be
\frac{\left(i\pi^{d/2}\right)^2 {\rm e}^{-2\gm_{\rm E}\ep}}{(-s)^{3+2\ep}}
\left[
\frac{4}{\ep^4} - \frac{9\pi^2}{2\ep^2} -\frac{53\zeta(3)}{3\ep}
+\frac{22\pi^4}{9}
- \pi i
\left(\frac{5}{\ep^3}-\frac{25\pi^2}{6\ep} -\frac{148\zeta(3)}{3}
\right)
\right] \; .
\ee

The presented algorithm is applicable to massless on-shell
box Feynman integrals with any integer powers of propagators.

\vspace{0.5 cm}

{\em Acknowledgments.}
I am grateful to M.~Beneke and the CERN Theory Group for kind
hospitality during my visit to CERN in April--May 1999 where
this work was completed. Thanks to M.B. for careful reading the
manuscript and to O.L.~Veretin for involving me into this problem
and useful discussions.
This work was supported by Volkswagen Foundation, contract
No.~I/73611, and by the Russian Foundation for Basic
Research, project 98--02--16981.


\begin{thebibliography}{99}

\bibitem{UD}
N.I. Ussyukina and A.I. Davydychev, {\em Phys. Lett.} B298 (1993) 363.

\bibitem{dimreg}
G.~'t Hooft and M.~Veltman, {\em Nucl.~Phys.} B44 (1972) 189;
C.G.~Bollini and J.J.~Giambiagi, {\em Nuovo Cim.} 12B (1972) 20.

\bibitem{Lewin}
L.~Lewin, {\em Polylogarithms and associated functions}
      (North Holland, 1981).

\bibitem{Devoto}
A.~Devoto and D.W.~ Duke, {\em Riv. Nuovo Cim.} 7 (1984) 1.

\bibitem{Oleg}
J. Fleischer, A.V. Kotikov and O.L. Veretin,
{\em Nucl.Phys.} B547 (1999) 343.

\end{thebibliography}
\end{document}